\documentclass[pre,reprint, showpacs,superscriptaddress]{revtex4-1} 

\usepackage{hyperref}
\usepackage{amsmath}
\usepackage{amsfonts}
\usepackage{pinlabel-test}
\usepackage{color}

\renewcommand{\d}[1]{\mathrm{d} #1}

\newcommand{\ds}{\displaystyle}

\newcommand{\eu}{\mathbf{e}_\theta}
\newcommand{\ed}{\mathbf{e}_\phi}
\newcommand{\n}{\mathbf{n}}
\newcommand{\tbf}{\mathbf{t}}
\newcommand{\nbf}{\boldsymbol{\nu}}

\newcommand{\R}{\mathbb{R}}
\newcommand{\ubf}{\mathbf{u}}
\newcommand{\vbf}{\mathbf{v}}

\newcommand{\dvol}{\d{A}}
\renewcommand{\div}{\text{div}\,}
\newcommand{\curl}{\text{curl}\,}
\newcommand{\divs}{\text{div}_s}
\newcommand{\curls}{\text{curl}_s}

\begin{document}

\title{Equilibrium configurations of nematic liquid crystals on a torus}

\author{Antonio Segatti} 
\address{Dipartimento di Matematica ``F. Casorati", Universit\`a di 
Pavia, Pavia, Italy} 
\author{Michael Snarski} 
\address{Brown University, Providence, USA} 
\author{Marco Veneroni} 
\address{Dipartimento di Matematica ``F. Casorati", Universit\`a di 
Pavia, Pavia, Italy}

\begin{abstract} 
The topology and the geometry of a surface play a fundamental role in 
determining the equilibrium configurations of thin films of liquid crystals. 
We propose here a theoretical analysis of a recently introduced surface 
Frank energy, in the case of two-dimensional nematic liquid crystals 
coating a toroidal particle. Our aim is to show how a different modeling 
of the effect of extrinsic curvature acts as a selection principle among 
equilibria of the classical energy, and how new configurations emerge. In 
particular, our analysis predicts the existence of new stable equilibria with 
complex windings. 
\end{abstract}


\maketitle


\section{Introduction}
 Due to their special optical properties and their controllability through 
electric and magnetic fields, liquid crystals have proven to be fundamental 
in many scientific and technological applications. Their properties have 
been deeply investigated for over a century and nowadays increasing 
emphasis is being placed  on so-called nematic shells. These are 
microscopic particles coated by a thin film of liquid crystals, which 
develop defects with a topological charge, and thus have a tendency to 
self-assemble into metamaterials which may have new optical properties 
and a high potential for technological applications (see, e.g., \cite
{Nelson02,YiEtal13}). The form of the elastic energy for nematics is well 
established, both in the framework of director theory, based on the works 
of Oseen, Zocher, and Frank, and in the framework of the order-tensor 
theory introduced by de Gennes (see, e.g., \cite
{Virga94,DegPro93,SegFri2}). In contrast, there is no universal agreement 
on the \emph{two-dimensional} free energy to model nematic shells. 
Different ways to take into account the distorsion effect of the substrate 
were proposed in \cite{Straley71,HelPro88,LubPro92} and recently by 
Napoli and Vergori (\cite{NapVer12E,NapVer12L}). Indeed, as 
observed in \cite{VitNel06} and \cite{BowGio2009}, the liquid crystal 
ground state (and all its stable configurations, in general) is the result of 
the competition between two driving principles: on one hand the 
minimization of the ``curvature of the texture" penalized by the elastic 
energy, and on the other the frustration due to constraints of geometrical 
and topological nature, imposed by anchoring the nematic to the surface 
of the underlying particle. The new energy model (\cite
{NapVer12E,NapVer12L}) affects these two aspects, focusing on the 
effects of the extrinsic geometry of the substrate on the elastic energy of 
the nematics. 
It is worthwhile noting that the above-mentioned
energies are deduced by means of heuristic considerations
or  ad hoc Ans\"atze. A rigorous justification, which
could indicate which form of energy is preferable, is still missing.
For instance, one could envision to model molecular interactions on a
discrete lattice and derive a macroscopic surface energy via $\Gamma$-convergence,
as it is done for the flat case in \cite{BCSarxiv}.

With the present paper we aim at exploring the full 
consequences of the new model so that a detailed comparison with the 
classical one can be established. More precisely, we study the two-dimensional
Napoli-Vergori director theory for nematic shells on a genus 
one surface: \emph{a)} we analyze the dependence of the new energy 
on the mechanical parameters (splay, twist and bend moduli) and on the 
geometrical parameters (the radii of the torus); \emph{b)} we highlight in 
which cases the new energy acts as a \emph{selection principle} among 
the minimizers of the classical one, and in which cases \emph{new} 
different states emerge. 
Finally,
\emph{c)} we predict the existence of stable 
equilibrium states carrying a higher energy than the ground state, in 
correspondence with the homotopy classes of the torus. Our analysis, in 
particular, agrees and makes more precise the statement of \cite
{NapVer12L}, according to which the new energy  ``promotes the 
alignment of the flux lines of the nematic director towards geodesics and/
or lines of curvature of the surface". The aspect of high energy 
equilibrium states is present in the classical energy as well, but it was 
neglected in previous research on genus one surfaces (\cite{LubPro92}).

Our observations are based on a rigorous mathematical analysis of the 
models, which combines methods from differential geometry and 
topology, calculus of variations, functional analysis and numerical 
simulations. Topology enters our work, first of all, in the choice of the torus 
as base substrate. Indeed the nematics would necessarily present defects 
on any surface with genus different than one, due to Poincar\'e-Hopf 
Theorem (\cite{Nelson83,MacLub91,LubPro92}); as a consequence, 
when dealing with the Frank's director theory, the space of functions in 
which one looks for minimizers would be empty, even in a weak sense 
(see \cite{CaSeVe}), requiring thus further expedients or approximations 
(see, e.g., \cite{KRV11,VitNel06}). The idea in \cite{CaSeVe} relies
on the extension of classical degree theory from continuous functions to
the space of Bounded Mean Oscillation functions, which, in particular,
 includes all the vector fields that have bounded Frank-Oseen energy \cite{BN1}. 
The emptiness of the set of minimizers of Frank-Oseen energy follows then from the emptiness of the set of continuous unit tangent vectors on surfaces with genus $\neq 1$. 
In order to focus on the influence of 
the geometry, we choose here a surface where defectless ground states can be 
found.  The case of a cylinder, the simplest surface where different results 
between classical and new energy can be predicted, was presented in 
\cite{NapVer12L} (on a sphere, the two energies differ by a constant). A 
related energy on hyperbolic surfaces was studied in \cite{Giomi12} and 
in \cite{Santangelo2012}, which described the different effects of the intrinsic 
and of the extrinsic geometry on defects. 
Although the experimental generation of toroidal nematics is a challenge, 
recent techniques (\cite{Pairam2013}) allow for droplets of genus one or 
higher, and further motivate investigation of more complex surfaces. 

\section{Energetics}
For vectors $\ubf,\vbf \in \R^3$ we denote by $\ubf \cdot \vbf$ the standard euclidean inner product. We write $| \ubf |  = \sqrt{\ubf \cdot \ubf}$ to denote
the norm of $\ubf$.  For a two-tensor $\mathbb A=\{a_{i}^j\}$ we adopt the norm $|\mathbb A |^2:= \text{tr}(\mathbb A^T \mathbb A)=\sum_{ij}(a_{i}^j)^2$, which is invariant under change of coordinates.

In the classical director theory of nematics, the local orientation of the  
liquid crystal molecules in a sample $\Omega\subset \R^3$ is described 
by the unit vector field $\n:\Omega \to \mathbb{S}^2$, where $\mathbb
{S}^2$ is the unit sphere.  Stable configurations are minimizers of the 
classical elastic energy, which according to Frank's formula reads
\begin{align*}
	W(\n)&:=\frac 12 \int_\Omega \big[K_1 (\div \n)^2 + K_2 (\n \cdot 
\curl \n)^2\\
		&  \quad +K_3 |\n \times \curl \n|^2 \\
		& \quad + (K_2 + K_{24})\div[(\nabla \n) \n -(\div \n)\n]
\big]\d x,
\end{align*}
 where $K_1$, $K_2$,  $K_3$  and $K_{24}$ 
are positive constants called the splay, twist, bend and saddle-splay 
moduli, 
 respectively. The last term is a null Lagrangian, hence
 it depends only on the behavior of $\n$ on the boundary. 
 While the Frank's energy above is, within the director theories, well 
accepted, 
 there is not such agreement when dealing with surface energies. In the 
literature,
 one can find different proposals for such an energy (\cite
{Straley71,LubPro92,VitNel06}
 \cite{NapVer12L}). The main difference between
 the classical energy proposed in \cite{Straley71,LubPro92,VitNel06} and 
the most recent one
 \cite{NapVer12L} essentially lies in the choice of 
 the differential operators on the surface $S$. More precisely, the 
 energy in (\cite{Straley71,LubPro92,VitNel06}) is a functional 
  of the covariant derivative $D\n$ of the vector field $\n$, 
 while the energy in \cite{NapVer12L} depends on the surface gradient
 $\nabla_{s}\n$, which is defined as
 $\nabla_{s}\n:=\nabla \n P$ (see, e.g., \cite{Virga94}), with $P$ being 
the 
 orthogonal projection onto the tangent plane of $S$. 
  Note that  
 \begin{equation}
 \label{expl}
 	D\n = P\nabla\n \neq\nabla\n P =\nabla_s\n,
 \end{equation}
	since the matrix product is non-commutative in general.
 In other words, $
\nabla_{s}$
 is the restriction of the usual derivative of $\mathbb{R}^3$ to directions 
lying in the tangent
 plane and takes thus into account also the extrinsic curvature of $S$. 
 In order to understand the effect of extrinsic curvature on $\nabla_s$ and the precise difference between the terms in \eqref{expl}, it is useful to introduce the 
 shape operator of a surface.  
 Denote by $\nbf$ the unit normal vector to $S$. 
At every point $p\in S$, $-\d \nbf_p:T_p S \to T_p \mathbb{S}^2$ is a symmetric linear operator named the \emph{shape operator}, the eigenvalues of which are exactly the two principal curvatures of $S$ in $p$.
 Given a smooth tangent vector field $\ubf$ on $S$, we can decompose the vector $(\nabla_s\n)\ubf$ into its projection on the tangent space and a normal component; by Gauss' formula \cite[Theorem 8.2]{lee}  we can  relate the normal component to the shape operator:
 \[ 
 	(\nabla_s\n)\ubf = P(\nabla_s\n)\ubf + [(\nabla_s\n)\ubf]^\perp = D_\ubf \n - (\d \nbf(\n)\cdot \ubf)\nbf.
\]	
The inner product in the last term is also known as the second fundamental form of $S$ and is usually denoted by $II(\n,\ubf)$.  This decomposition implies the  identity
\begin{equation}
\label{expl2}
	|\nabla_s \n|^2 = |D\n|^2 +|\d\nbf(\n)|^2,
\end{equation} 
 which clarifies the difference between covariant derivative and surface gradient.
 In order to describe the elastic energy of a thin film, approximated by a 
surface $S$, we resort to the Darboux frame $\{\n,\tbf,\nbf\}$, where 
 $\tbf:=\nbf \times \n$. Let $\kappa_\tbf,\kappa_\n$ be the geodesic curvatures of the flux lines of $
\tbf$ and $\n$, respectively. Let $c_\n$ be the normal curvature and let 
$\tau_\n$ be the geodesic torsion (see, e.g., \cite{DoCarmo76} for all 
the definitions and examples related to differential geometry). The surface 
divergence
   acting on the field $\n$ is defined as $\divs\n:=\hbox{tr}D\n =
   \hbox{tr}\nabla_{s}\n$ and can be expressed as 
   $\divs \n=\kappa_\tbf$
   (\cite{NapVer12L}). 
 The classical form of surface free energy, for a thin film of nematics of 
constant thickness $h$ around $S$, is (\cite
{Straley71,LubPro92,VitNel06})                  
\begin{align}
	W_\text{Cl}(\n):&=\frac 12 \int_S \left[k_1 (\divs \n)^2 + k_3 
	(\curl \n)^2\right]\dvol, \nonumber\\
		&= \frac 12 \int_S \left[k_1 \kappa_\tbf^2 + k_3 \kappa_
\n^2\right]\dvol,
\end{align}
where $k_i=h K_i$ and $\curl\n$ is the covariant 
curl operator (see \cite{lee}).
In the new energy introduced in \cite{NapVer12L}, the
extrinsic curvature of $S$ comes to play a role through $c_\n$ and $
\tau_\n$  
\begin{align}
	W_\text{NV}(\n):&=\frac 12 \int_S \left[k_1 (\divs \n)^2 + k_2(\n 
\cdot \curls \n)^2\right. \nonumber\\
	&\quad \left.+k_3 |\n \times \curls \n|^2\right]\dvol,\label{eq:ennv}\\
		&= \frac 12 \int_S \left[k_1 \kappa_\tbf^2 + k_2 \tau_\n^2 
	  +k_3 (\kappa_\n^2+c_\n^2)\right]\dvol,\nonumber
\end{align} 
where $\curls\n:=-\epsilon\nabla_{s}\n$ ($\epsilon$ is the Ricci 
alternator).
Note that
$\curls \n = -\tau_\n \n - c_\n \tbf + \kappa_\n \nbf$
and that, unless we restrict to flat surfaces, the vector $\curl_s \n$ 
has non vanishing in-plane components. Note also that the saddle-splay 
term
is not present in the surface energy $W_\text{NV}$ (see \cite{NapVer12L} for 
a justification). 
In order to study the minimizers of $W_\text{Cl}$ and $W_\text{NV}$, it 
is convenient to introduce a parametrization $X$ of $S$. We use $
(\theta,\phi)$ as a set of local coordinates and we assume that $\eu:=
\partial_\theta X /|\partial_\theta X |$ and $\ed:=\partial_\phi X /|
\partial_\phi X |$ give a local orthonormal basis for the tangent plane to 
$S$. We can then describe $\n$ through the angle $\alpha$ defined by 
$\n=\eu \cos \alpha +\ed \sin \alpha$. With respect to $\alpha$, the 
surface energy \eqref{eq:ennv} takes the form
\begin{align} 
 	W_\text{NV}(\alpha) =& \frac 12 \int_S \left[k_1 ((\nabla_s \alpha -
\boldsymbol{\Omega})\cdot \tbf)^2 \right. \nonumber\\
		&+ k_3 ((\nabla_s \alpha -\boldsymbol{\Omega})\cdot \n)^2\nonumber\\
		&+ k_2(c_1-c_2)^2\sin^2\alpha\,\cos^2\alpha \nonumber\\
		& \left.+k_3(c_1\cos^2\alpha+ c_2\sin^2\alpha )^2\, \right] \dvol, \label{eq:enealpa_gen}
	\end{align}	
where $c_1$ and $c_2$ are the principal curvatures of $S$ and $
\boldsymbol{\Omega}$ is the spin connection. The latter is the vector field defined as
$ \boldsymbol{\Omega} = -\kappa_\theta \eu -\kappa_{\phi}\ed$,
where $\kappa_\theta$ and $\kappa_\phi$ are the
geodesic curvatures of the flux lines of
$\eu$ and $\ed$, respectively
 (see \cite{NelPel87} and the Appendix \ref{app:torus} in this paper). The 
first two terms coincide with $W_\text{Cl}$.

\section{The Axisymmetric Torus}
In this Section we analyse the energy \eqref{eq:enealpa_gen} in the particular 
case of an axisymmetric torus $\mathbb T$, with radii $R>r>0$ 
parametrized by 
\begin{equation}
\label{eq:paramtorus}
	X(\theta,\phi) = 
		\begin{pmatrix} 
			(R+r\cos \theta)\cos \phi \\ 
			(R+ r\cos \theta)\sin \phi \\ 
			r\sin \theta
		\end{pmatrix},
\end{equation}
\begin{figure}[h!]
\begin{center}
\labellist
		\hair 2pt
		\pinlabel $\phi$ at 265 200
		\pinlabel $\mathbb{T}$ at 150 320	
		\pinlabel $\theta$ at 360 200
		\pinlabel $R$ at 295 165		
		\pinlabel $r$ at 330 220				
	\endlabellist				
\centering{
\includegraphics[height=4.8cm]{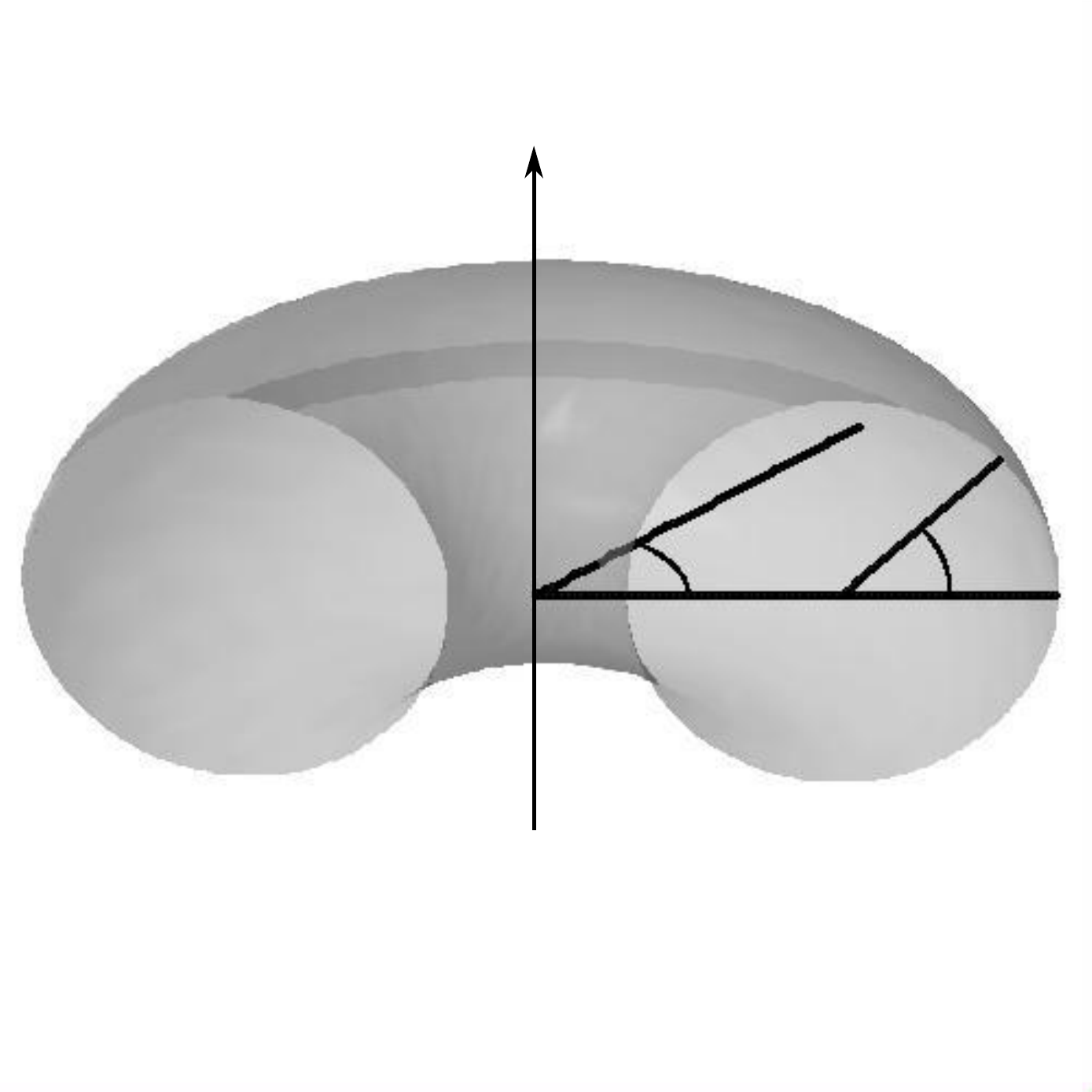}\\  \vspace{-1.5cm}
}
\end{center}
\caption{Schematic representation of the torus $\mathbb T$.}
\label{fig1}
\end{figure}
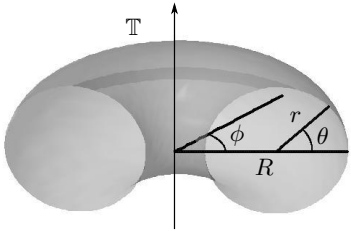
where $(\theta,\phi)\in Q:=[0,2\pi]\times [0,2\pi]$ (see Figure \ref{fig1}). 
In order to study the 
dependence of the energy \eqref{eq:ennv} on the mechanical and 
geometrical parameters, we first restrict to the case of \emph{constant 
angle} $\alpha$. In this case, the integral in \eqref{eq:ennv} can be 
computed explicitly as a function of five real parameters (see Appendix B): 
\begin{equation}
\label{eq:Walpha}
	W_\text{NV} =W_\text{NV}(\alpha,k_1,k_2,k_3,\mu),
\end{equation}	
where $\mu:=R/r$. 
\begin{figure}[h!]
\begin{center}
\labellist
		\hair 2pt
		\pinlabel $\alpha_m$ at 15 400
		\pinlabel $\alpha_p$ at -30 200	
		\pinlabel $\alpha_h$ at 605 200
	\endlabellist				
\centering{
\includegraphics[height=5cm]{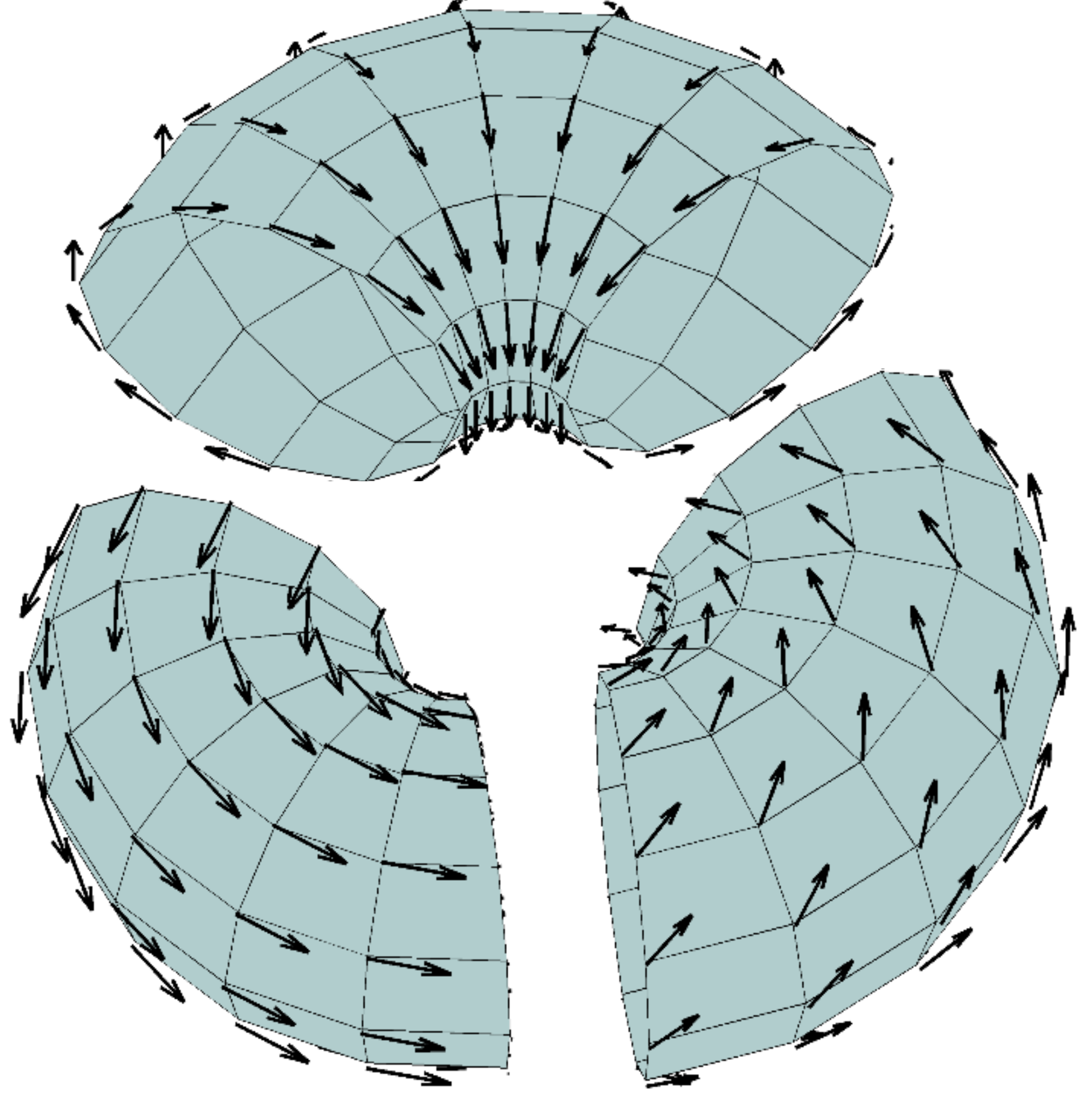}\\  \vspace{-0.6cm}
}
\end{center}
\caption{Schematic representation of the constant equilibrium 
configurations $\alpha_m$, $\alpha_p$ and $\alpha_h$, where the 
director field $\n$ is aligned along meridians,  parallels and helixes of $
\mathbb T$, respectively.}
\label{fig2}
\end{figure}
Since $W_\text{NV}$ is $\pi$-periodic, we restrict to $\alpha \in ]-\pi/
2,\pi/2]$.
This choice is related to the invariance of the energy $W_\text{NV}$ 
with respect to the change of $\n$ into its opposite $-\n$.
 Studying the equilibrium equation associated with $W_\text
{NV}$, one finds the equilibrium configurations (see Figure \ref{fig2}) 
\[
 	\alpha_m = 0,\quad \alpha_p = \frac \pi 2,\quad
 \alpha_h=\pm\frac 12\arccos\left(\frac{Bk_3 +Ck_1}{\mu^2(k_2-k_3)}
\right),
\]
where $B=\mu\sqrt{\mu^2-1}-1$ and $C=B-\mu^2+2$ (provided that the 
argument of the arccos function is in the interval $[-1,1]$).

\begin{figure}[h]
		\labellist
			\hair 2pt
			\pinlabel $\ds\frac{W_\text{NV}}{k\pi^2}$ at 220 307
			\pinlabel $\alpha$ at 384 -4
			\pinlabel $-\frac{\pi}{2}$ at 30 -4
			\pinlabel $0$ at 196 -4									
			\pinlabel $\frac{\pi}{2}$ at 350 -4
			\pinlabel $2$ at 180 70									
			\pinlabel $4$ at 180 270			
		\endlabellist				
		\includegraphics[height=6cm]{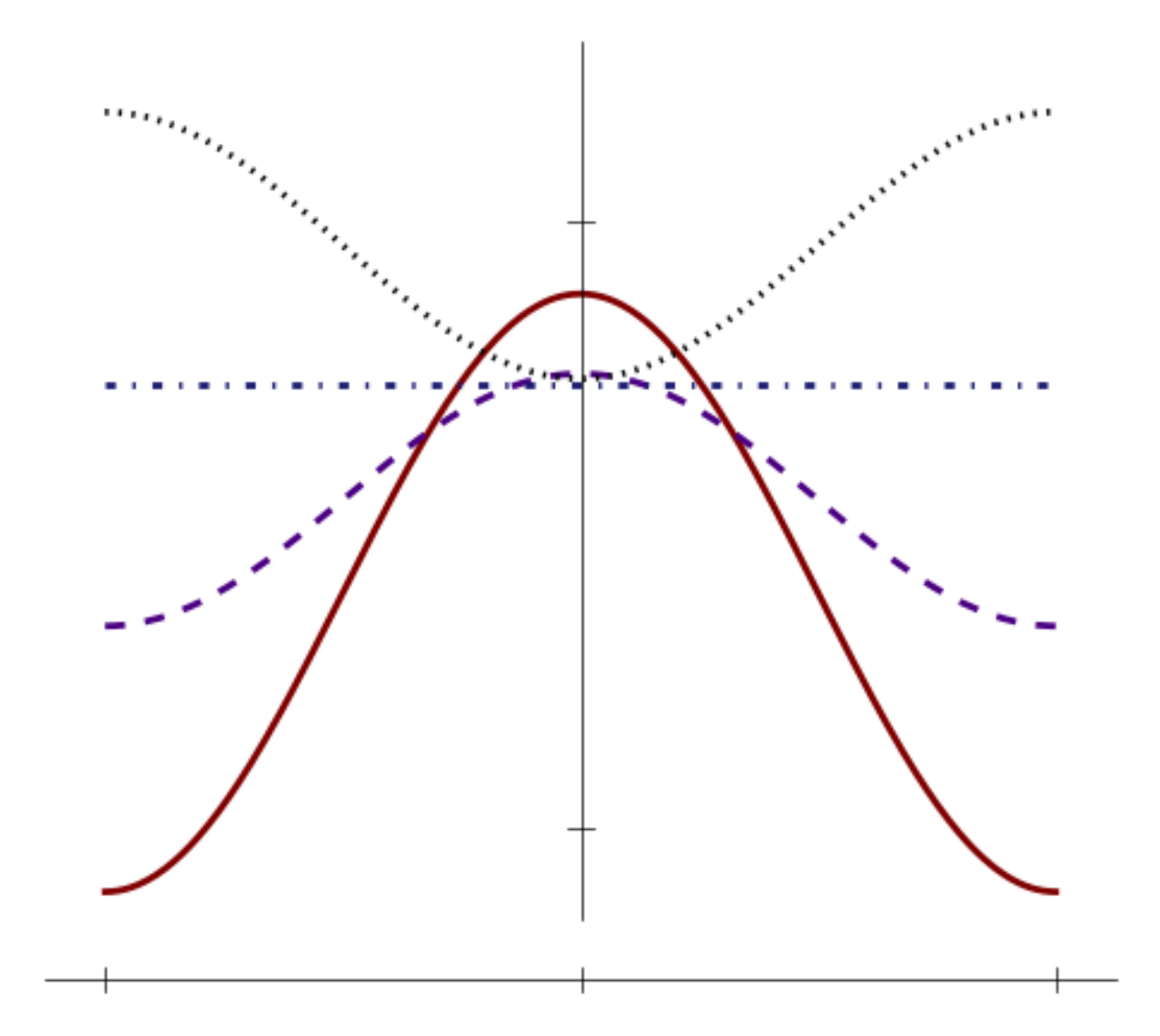}
		\caption{Surface energy $W_\text{NV}$ (rescaled by $k\pi^2$) as a 
function of the deviation $\alpha$ from $\eu$, for the one-constant 
approximation $k:=k_1=k_2=k_3$. The ratio of the radii of the torus $
\mu=R/r$ is : $\mu=1.1$ (dotted line), $\mu= 2/\sqrt 3$ (dashed dotted 
line), $\mu=1.25$ (dashed line), $\mu=1.6$ (continuous line). The 
behavior for equal $k_j$'s depends on the fact that the only non-constant 
term in the energy $W_\text{NV}$ changes sign when $2\mu=
\mu^2/\sqrt{\mu^2-1}$, i.e. when $\mu=2/\sqrt 3$. } 
	\label{fig3}
\end{figure}
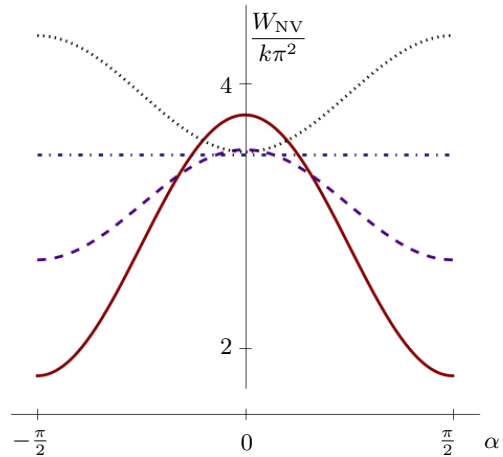

\begin{figure}[h]
		\labellist
			\hair 2pt
			\pinlabel $\ds\frac{W_\text{NV}}{k_1\pi^2}$ at 220 215
			\pinlabel $\alpha$ at 384 -4
			\pinlabel $-\frac{\pi}{2}$ at 30 -4
			\pinlabel $0$ at 196 -4									
			\pinlabel $\frac{\pi}{2}$ at 350 -4
			\pinlabel $2$ at 180 119									
			\pinlabel $4$ at 180 218			
		\endlabellist				
		\includegraphics[height=5cm]{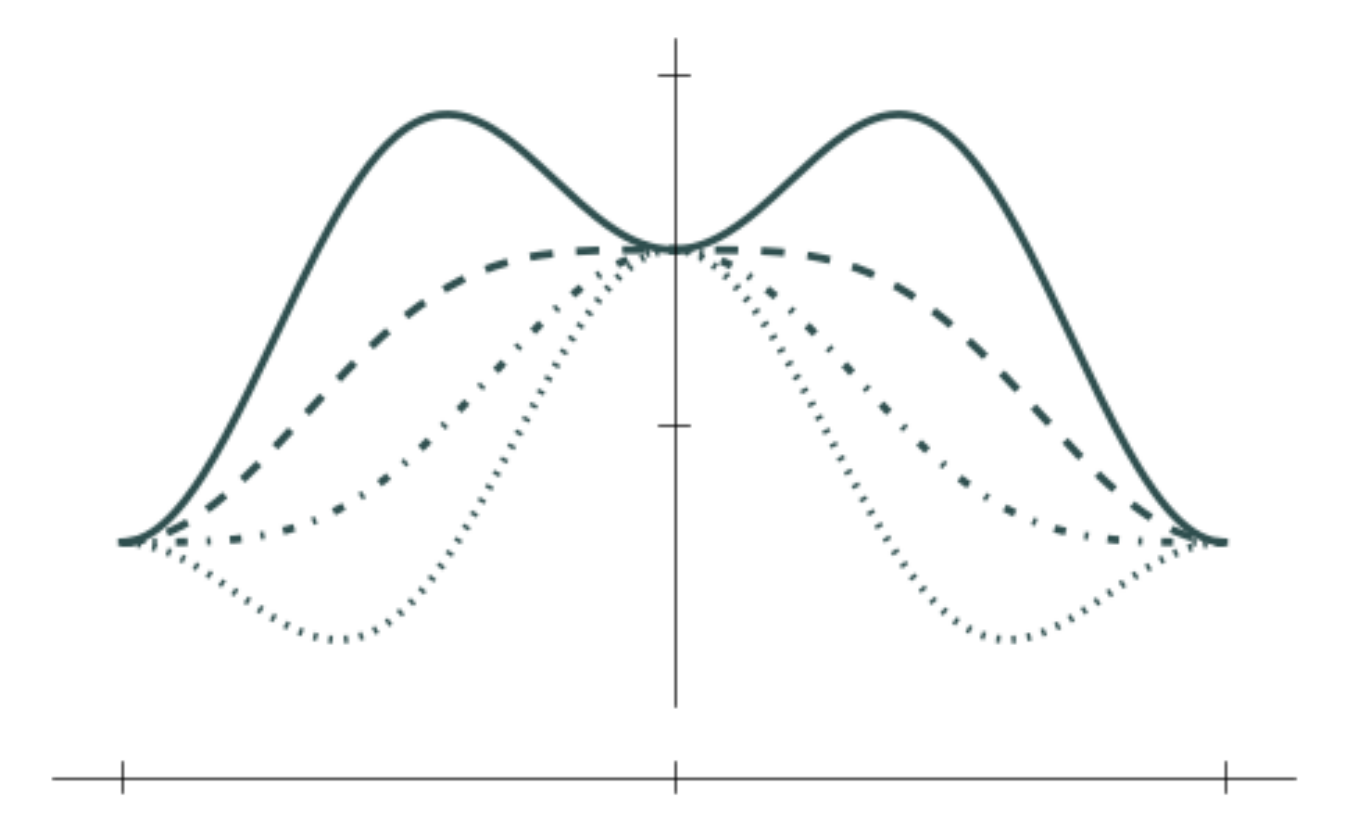}
		\caption{Surface energy $W_\text{NV}$ (rescaled by $k_1\pi^2$) 
as a function of the deviation $\alpha$ from $\eu$, for $\mu=1.25$,  
$k_1=k_3=1$ and different values of $k_2$. Studying the second 
derivative of $W_\text{NV}$, one finds critical values $\xi_1=2\sqrt
{\mu^2-1}/\mu$ and  $\xi_2=2-\xi_1$, which characterize the cases:  
		$\alpha_m$ is local minimum and $\alpha_p$ is global minimum 
($k_2>\xi_1$, continuous line); 
		$\alpha_m$ is global maximum and $\alpha_p$ is global minimum 
($\xi_1\geq k_2 \geq \xi_2$, between dashed and dashed-dotted lines); 
		$\pm \alpha_h$ are global minima, $\alpha_p$ is local maximum, 
$\alpha_m$ is global maximum ($k_2<\xi_2$, dotted line).}
	\label{fig4}
\end{figure}
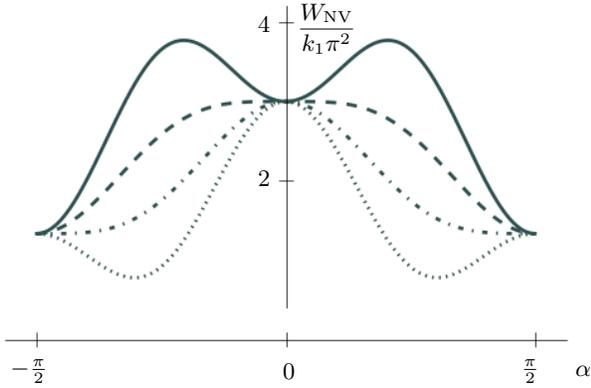

Unlike the case of a cylinder \cite{NapVer12L} in which the radius has
no influence on minimizers, in the case of a torus the minimality 
depends also on the ratio $\mu=R/r$. In this regard, see Figure \ref
{fig3}, where $W_\text{NV}(\alpha)$ is plotted for fixed $k_j$ and 
different choices of $\mu$, and Figure \ref{fig4}, which shows $W_\text
{NV}(\alpha)$ for fixed ratio $\mu$ and different choices of $k_j$.

In Figures \ref{fig3} and \ref{fig4} we make a specific choice of the parameters $k_i$ and $\mu$ so as to present the features of the energy. The advantage of the explicit formula for $W_\text{NV}$ given in Appendix B is that we may assess the minimality and stability of a configuration for any given choice of parameters simply by computing the first and second derivatives of $W_\text{NV}$ with respect to $\alpha$.

We turn now our attention to \emph{general} functions $\alpha$ (i.e., not 
necessarily constants), in the case of the well-studied one-constant 
approximation of $W_\text{NV}$, where $k_1=k_2=k_3=:k$. Let $
\nabla_{s} \n$ be the surface derivative of $\n$ and let 
$\nbf$ be the unit normal vector to the torus $\mathbb T$.
Recalling that (see \cite{NapVer12L})
\[
(\divs \n)^2 + (\n \cdot \curls \n)^2 + |\n \times \curls \n|^2 = |\nabla_s \n|^2,
\]
 we can easily obtain the one-constant approximation of $W_\text{NV}$ 
\begin{align*}
	W_\text{NV}(\n)&=\frac{k}{2}\int_{\mathbb T} |\nabla_{s} \n|^2 \dvol\\
			& \stackrel{\eqref{expl2}}{=}\frac{k}{2}\int_{\mathbb T} \left[|D\n|^2 +|\d\nbf(\n)|^2\right] \dvol.	
\end{align*}
 It is then clear that by introducing the surface gradient $\nabla_s$ (with respect to the tangent derivation), one is adding a contribution which depends on the extrinsic curvatures. More precisely, $|\d\nbf(\n)|$ is minimized (maximized) when $\n$ is oriented along the direction of minimal (maximal) curvature (in absolute value).  We can check this directly in the case of a torus by expressing
 the energy 
in terms of the angle $\alpha$ with the local coordinates,
\begin{align}
\label{eq:enalpa}
	W_\text{NV}(\alpha) = f(\mu)+\frac{k}{2}\int_{\mathbb T} \left[|
\nabla_s \alpha|^2 +\frac{c_1^2-c_2^2}{2}\cos(2\alpha)\right]\dvol,
\end{align}
where $f(\mu)=\kappa \pi^2 (2\mu + (2-\mu^2)/\sqrt{\mu^2-1}))$ can 
be computed using the orthogonality of $\nabla_s\alpha$ and $
\boldsymbol{\Omega}$.

\begin{figure}[h]
		\labellist
			\hair 2pt
			\pinlabel $\alpha_\infty$ at 1775 720
		\endlabellist				
		\includegraphics[height=5cm]{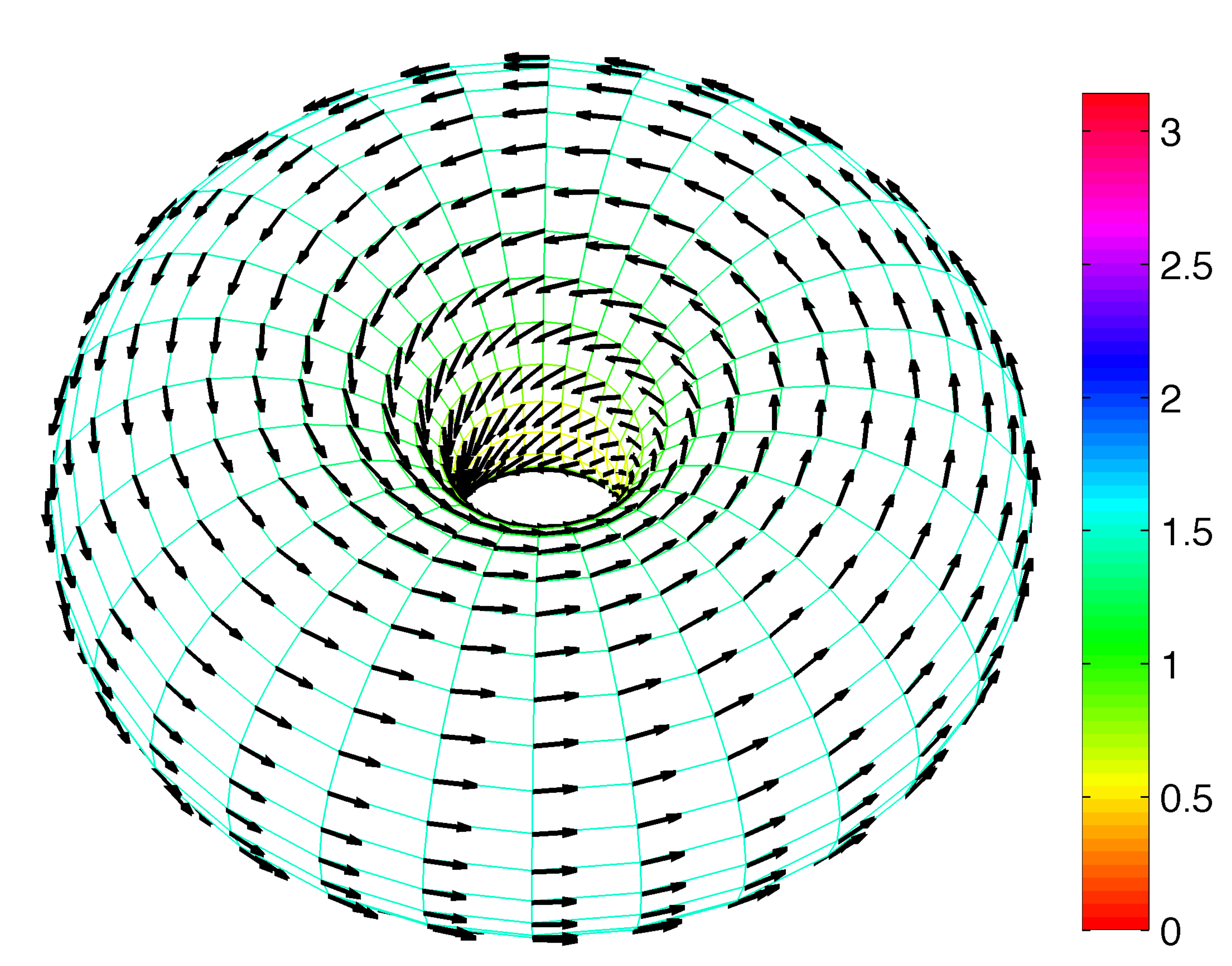}\\
		$a)$ $\mu=1.4$\\
		\medskip
		
		\includegraphics[height=5cm]{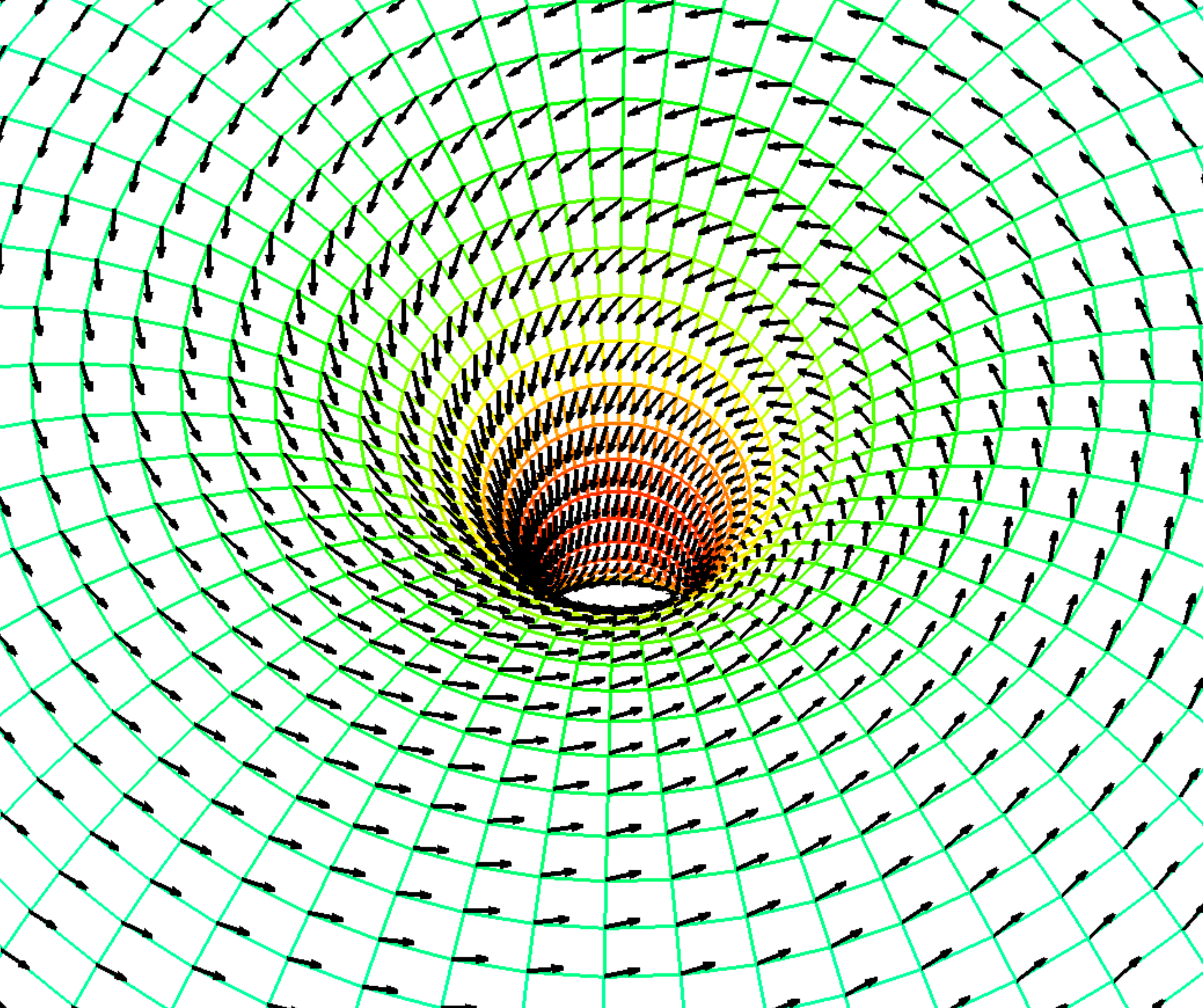}\\
		$b)$ $\mu=1.2$\\
		\caption{(Color online) Plots of the vector field $\n_\infty$ corresponding to the limit 
solution $\alpha_\infty$ to \eqref{eq:gf} for $\alpha_0=\pi/4$ and 
different ratios $\mu$. (The color code represents $\alpha_\infty$.) For $
\mu<2$, the configuration $\alpha_m$ is preferable in a strip close to the 
central hole, while for every choice of $\mu$, $\alpha_p$ is more 
convenient around the external equator. As the term $|\nabla_s \alpha|
^2$ in the energy penalizes the transition from $\alpha_p$ to $\alpha_m
$, minimizers exhibit a smooth rotation of the director field along the 
meridians. The amplitude of the rotation increases as $\mu$ decreases to 
1.}
	\label{fig5}
\end{figure}

The energy $W_\text{NV}$ in the one-constant 
approximation consists in a Dirichlet energy density $|
\nabla_s \alpha|^2$ plus a double-(mod $2\pi$)-well potential $\cos(2\alpha)$ multiplied by a curvature-dependent factor $(c_1^2-c_2^2)/2$.
This form of the energy is well-studied in the context of Cahn-Hilliard phase transitions (see
\cite{cahn_hill}), where pure phases correspond to minimizers of the double-well and transitions are penalized by the Dirichlet term. 
In our case the liquid crystal phase is clearly fixed, but we can apply the same structure to a ``direction transition". Depending on the torus aspect ratio, 
the sign of $c_1^2-c_2^2$ may not be constant on $Q$, thus forcing 
a smooth transition between the directions $\alpha \equiv \alpha_m$, where $c_1^2<c_2^2$, 
and $\alpha \equiv \alpha_p $, where $c_1^2>c_2^2$ (see Figure \ref{fig5}).

The Euler-Lagrange equation corresponding to \eqref{eq:enalpa} is
\begin{equation}
\label{eq:el}
	k\Delta_s \alpha +\frac k2(c_1^2-c_2^2)\sin (2\alpha)=0,
\end{equation}
where $\Delta_s=\divs\nabla_s$ is the Laplace-Beltrami operator on the 
torus $\mathbb T$. Equation \eqref{eq:el} is a novel kind of elliptic sine-
Gordon equation, the only explicit solutions to which, to our knowledge, 
are the constants $\alpha_m=0$ and $\alpha_p=\pi/2$. We resort thus 
to studying the gradient flow of $W_\text{NV}$, i.e., the solutions $
\alpha=\alpha(x,t)$ defined on $\mathbb T \times [0,+\infty)$, to the 
evolution equation
\begin{equation}
\label{eq:gf}
	\partial_t \alpha = k\Delta_s \alpha +\frac k2(c_1^2-c_2^2)\sin 
(2\alpha),
\end{equation}
equipped with an initial datum $\alpha(x,0)=\alpha_0(x)$ on $\mathbb 
T$.  We remark that this evolution problem is not a physical flow of the 
nematics, but it constitutes an efficient mathematical artifice to 
approximate solutions of the stationary equation \eqref{eq:el}. Indeed, 
owing to the ellipticity of the Laplace-Beltrami operator and to the 
regularity of the nonlinear term, for any regular initial datum $\alpha_0$ 
there exists a unique solution $\alpha(t)$ to \eqref{eq:gf}. Moreover, by 
construction, at any time $t>0$ this solution satisfies the energy balance
\begin{equation}
\label{enbal}
	\int_0^t \int_{\mathbb T}|\partial_t \alpha(s)|^2\dvol\, \d s +W_\text
{NV}(\alpha(t))=W_\text{NV}(\alpha_0).
\end{equation}

The energy equality above follows from the fact that \eqref{eq:gf}
is indeed the gradient flow of the energy \eqref{eq:enalpa}. 
For a (sufficiently smooth) solution $\alpha$ of \eqref{eq:gf},
we have that 
\begin{align*}
	\frac{\d}{\d t}W_\text{NV}(\alpha(t)) 
		&=\frac{\d}{\d t}\frac{k}{2}\int_{\mathbb T} \Big[|\nabla_s \alpha(t)|^2 \\
		&\quad +\left.\frac{c_1^2-c_2^2}{2}\cos(2\alpha(t))\right]\dvol\\
		&= k\int_{\mathbb T}\Big[\nabla_s\alpha(t)\cdot\nabla_s\partial_t\alpha(t) \\
		&\quad -\left.\frac{c_1^2-c_2^2}{2}\sin(2\alpha(t))\partial_t\alpha(t) \right]\dvol.
\end{align*}
Integrating by parts, we obtain
\begin{align*}
	\frac{\d}{\d t}W_\text{NV}(\alpha(t)) 
		&= -k\int_{\mathbb{T}} \Big[\Delta_s\alpha(t)\\
		&\quad + \left.\frac{c_1^2-c_2^2}{2}\sin(2\alpha(t))\right]\partial_t\alpha(t)\dvol.
\end{align*}
Consequently, recalling that $\alpha$ is 
a solution of \eqref{eq:gf}, we get
\begin{equation}
\label{eq:dissipation}
	\frac{\d}{\d t}W_\text{NV}(\alpha(t)) = -\int_{\mathbb{T}}\vert \partial_t\alpha(t)\vert^2\dvol, \qquad\forall t>0,
\end{equation}
which expresses that the energy $W_\text{NV}$ decreases along
solutions. Integrating over $(0,t)$ for $t>0$ we obtain the energy balance
\eqref{enbal}.
Finally, as $t \to +\infty$, $\alpha(t)$ converges (possibly up 
to a subsequence) to a function $\alpha_\infty$ which solves \eqref
{eq:el}. If $\alpha_0$ is a critical point, i.e. $\alpha_0=\alpha_m$ or $
\alpha_0=\alpha_p$, then the evolution is clearly constant: $\alpha(t)
\equiv\alpha_m$, or $\alpha(t)\equiv\alpha_p$, respectively. 
For different initial data, we find as a limit solution either $\alpha_p$, or 
a nonconstant $\alpha_\infty$ belonging to the family illustrated in Figure 
\ref{fig5}. The only distinguishing factor between these two behaviors is 
the ratio of radii $\mu=R/r$. From the explicit form of $W_\text{NV}(\alpha)
$, we know that if $\mu \geq 2$, then $c_1^2-c_2^2\geq0$ and  thus $
\alpha_p\equiv \pi/2$ is the unique (up to rotations of $m\pi,m\in 
\mathbb Z$) global minimizer of $W_\text{NV}$. On the other hand, 
from the previous discussion on constant $\alpha$, we know that if $\mu 
< 2/\sqrt 3$, then $\alpha_p$ cannot be the global minimizer, as for this 
ratio $W_\text{NV}(\alpha_m)<W_\text{NV}(\alpha_p)$. 
We conjecture that there exists a unique critical ratio $\mu^* \in (2/\sqrt 
3,2)$ above which $\alpha_p$ is the point of minimum, and below which 
the nonconstant solution appears. Numerically, we found $\mu^*\approx 
1.52$. 

It is interesting to compare these configurations with the equilibrium ones 
of the classical Frank energy \cite{LubPro92}. In the one-constant 
approximation, the energy on a torus is
\begin{align*}
	W_\text{Cl}(\alpha) &=\frac{k}{2}\int_{\mathbb T}|\nabla_s \alpha -
\boldsymbol \Omega|^2\, \dvol\\
	&  =\frac{k}{2}\int_{\mathbb T}\left[|\nabla_s \alpha|^2 -2\nabla_s\alpha \cdot
\boldsymbol \Omega +|\boldsymbol\Omega|^2\right] \dvol\\
					&=\frac{k}{2}\int_{\mathbb T}|\nabla_s \alpha|^2\, \dvol + 
2k\pi^2(\mu-\sqrt{\mu^2-1})
\end{align*}
and the corresponding equilibrium equation is $\Delta_s \alpha=0$. 
(Note that in \cite{LubPro92}  $\mu=r/R$.) Therefore, in the classical 
case, every field $\n=\eu\cos \bar \alpha +\ed\sin \bar \alpha$, for 
constant $\bar \alpha$, is an equilibrium state, with the same energy 
independently of $\bar \alpha$. For $\mu>\mu^*$, the new energy 
$W_\text{NV}$ selects $\alpha_p$, among all constants, as unique 
equilibrium configuration. For $\mu<\mu^*$, instead, the new lower-energy 
configuration shown in Figure \ref{fig5} appears.
The fact that the solution $\alpha =\alpha_p$ is no longer stable for sufficiently small $\mu$ is due to the high bending energy associated to $\alpha =\alpha_p$ in the internal hole of the torus. In a small strip close to the internal equator of the torus, we can approximate (see Appendix B)
\[
	c_1^2 -c_2^2 \approx \frac{1}{r^2}-\frac{1}{(R-r)^2},\quad \d A \approx r(R-r)\d\theta\,\d\phi,
\] 
and therefore 
\begin{align*}
	(c_1^2-c_2^2)\cos(2\alpha_p)\dvol \approx \mu \frac{2-\mu}{\mu-1}\d\theta\,\d\phi,
\end{align*}
which tends to $+\infty$ as $\mu \to 1$.

This new solution attempts to minimize the effect of the curvature by
 orienting the director field along the meridian lines $\alpha = \alpha_m$ which
  are geodesics on the torus, near the hole of the torus, while near the
   external equator the director is oriented along the parallel lines $\alpha=\alpha_p$,
    which are lines of curvature. 
The particular form of the energy $W_\text{NV}$  in \eqref{eq:enalpa} favors a smooth transition 
between  $\alpha=\alpha_p$ and $\alpha=\alpha_m$.
This
``phase" transition is due to the interplay between the
Dirichlet term and the double well potential in the one-constant approximation
 \eqref{eq:enalpa}. The double-well potential is exactly 
the contribution of the extrinsic terms in the energy.

\subsection{Equilibria with windings}

In order to describe more complex equilibrium states, we need to 
introduce the winding number of the director field $\n$ on the torus. Let $
\n$ be given and, referring to \eqref{eq:paramtorus}, let $\alpha:Q\to 
\R$ be such that
\[
	\n(X)=\eu \cos \alpha +\ed \sin \alpha\quad \text{on }Q. 
\]
Though $\alpha$ needs not be $Q$-periodic, there certainly exist integers $m,n\in \mathbb Z$ such that
\begin{equation*}
	\alpha(2\pi,0)= \alpha(0,0) +m\pi,\qquad
	\alpha(0,2\pi)= \alpha(0,0) +n\pi.	
\end{equation*}
We define the \emph{winding number} of $\pm \n$ as the couple of indices $
\mathbf h =(h_\theta,h_\phi)\in \mathbb Z \times \mathbb Z$, given by
\begin{equation*}
	h_\theta:= \frac{\alpha(2\pi,0)- \alpha(0,0)}{\pi},\qquad
	h_\phi:=\frac{\alpha(0,2\pi) -\alpha(0,0)}{\pi}. 	
\end{equation*}

\begin{figure*}[ht]
		\labellist
			\hair 2pt
			\pinlabel $\mathbf{h}=(1,0)$,\quad$W=10.73$ at 850 -35
		\endlabellist				
		\includegraphics[height=3.5cm]{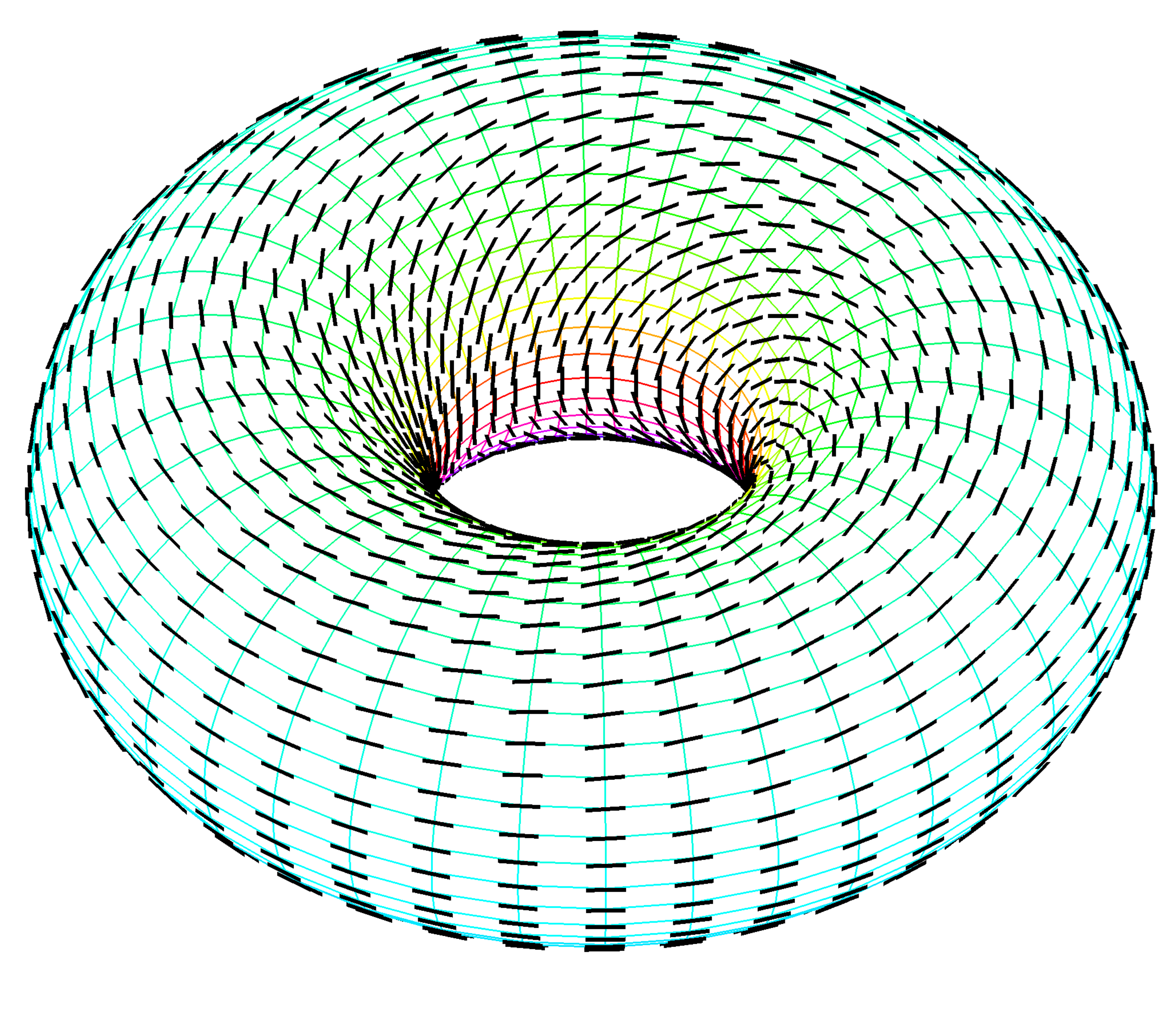} \qquad
				\labellist
			\hair 2pt
			\pinlabel $\mathbf{h}=(0,1),\quad\mbox{$W=10.93$}$ at 850 -35
		\endlabellist				
		 \includegraphics[height=3.5cm]{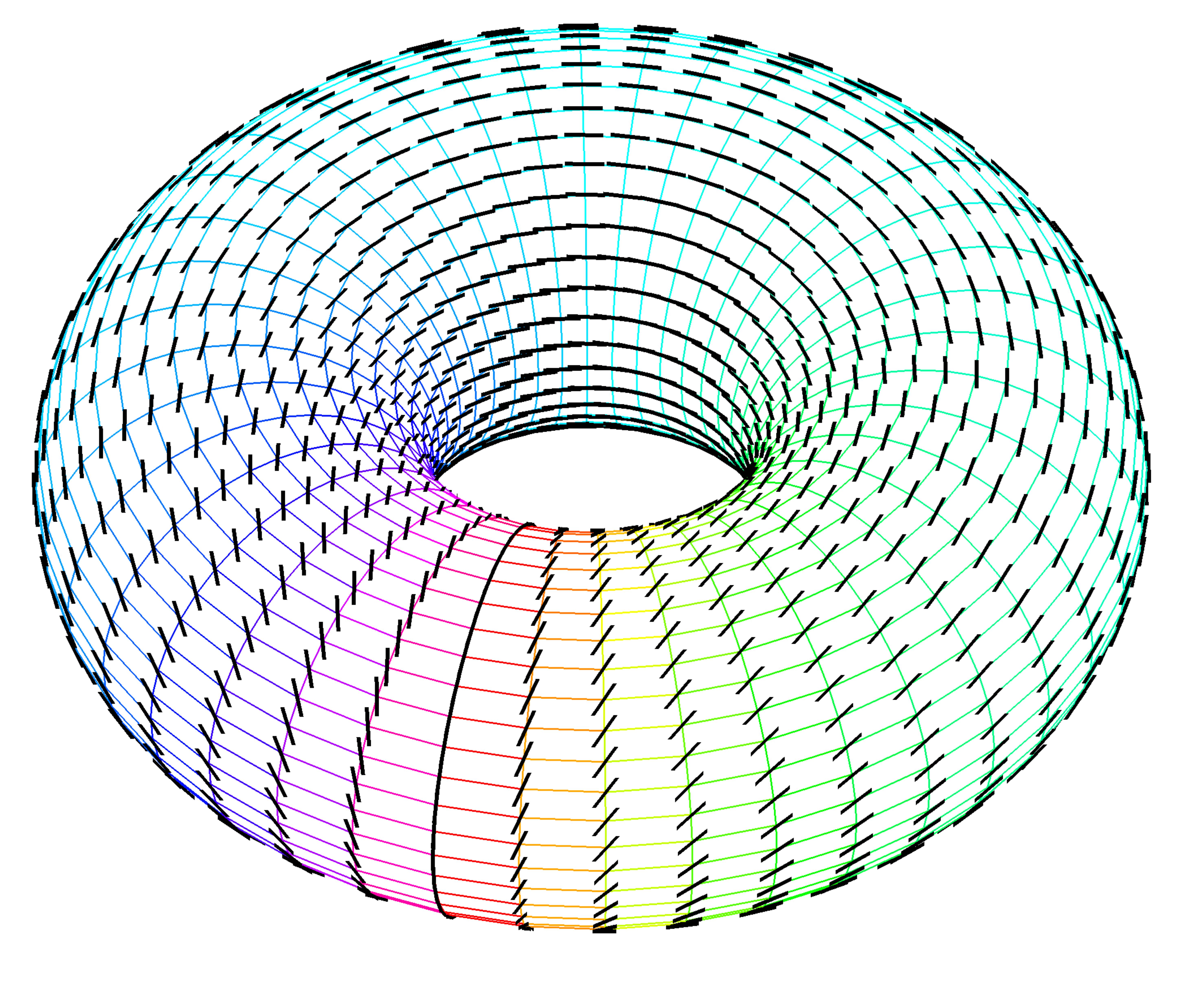} \qquad
		 \labellist
			\hair 2pt
			\pinlabel $\alpha$ at 1970 720
			\pinlabel $\mathbf{h}=(0,3),\quad\mbox{$W=14.01$}$ at 850 -35
		\endlabellist				
				\includegraphics[height=3.5cm]{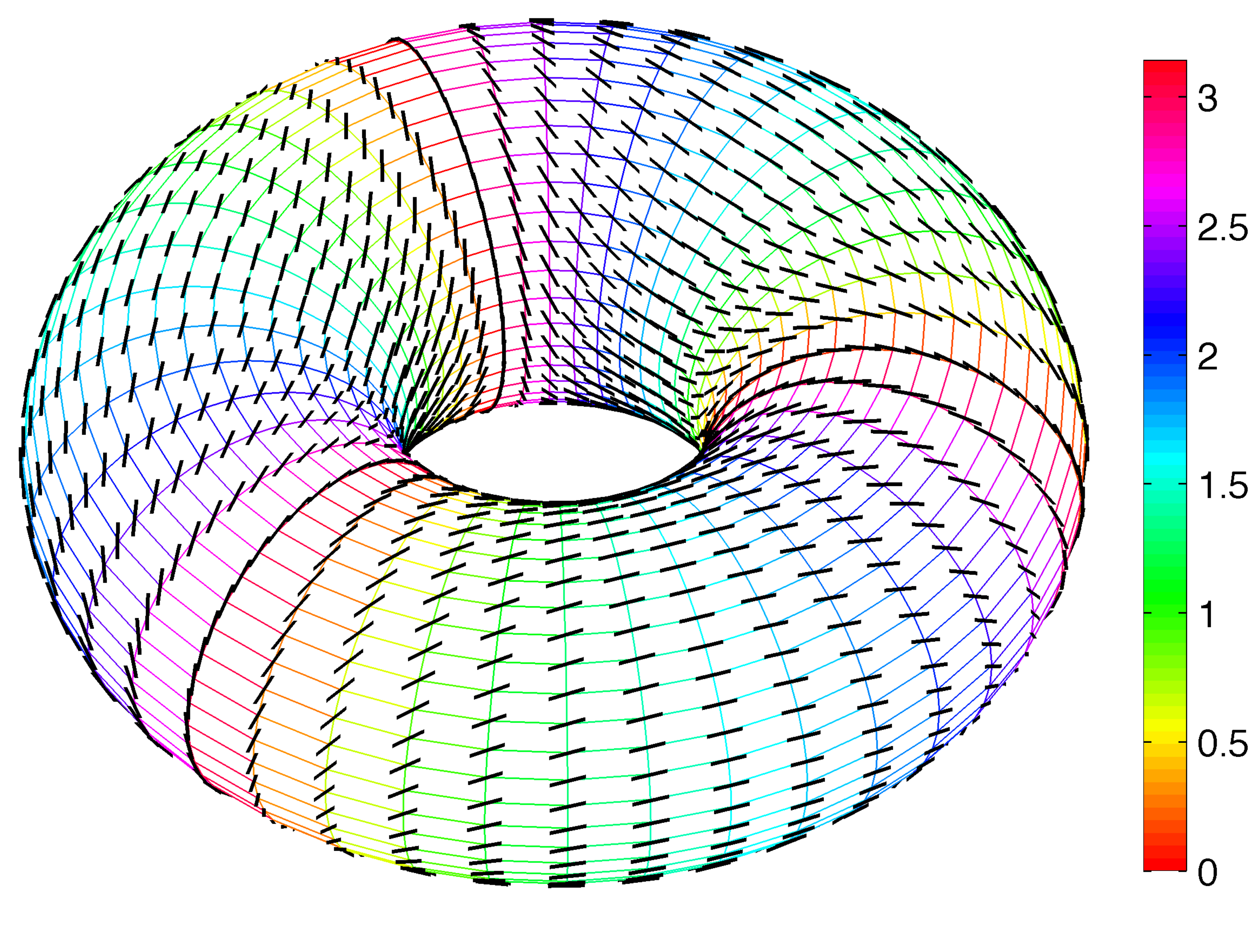} \\
				\bigskip
				
		\labellist
			\hair 2pt
			\pinlabel $\mathbf{h}=(1,1),\quad\mbox{$W=11.75$}$ at 850 -35
		\endlabellist				
		\includegraphics[height=3.5cm]{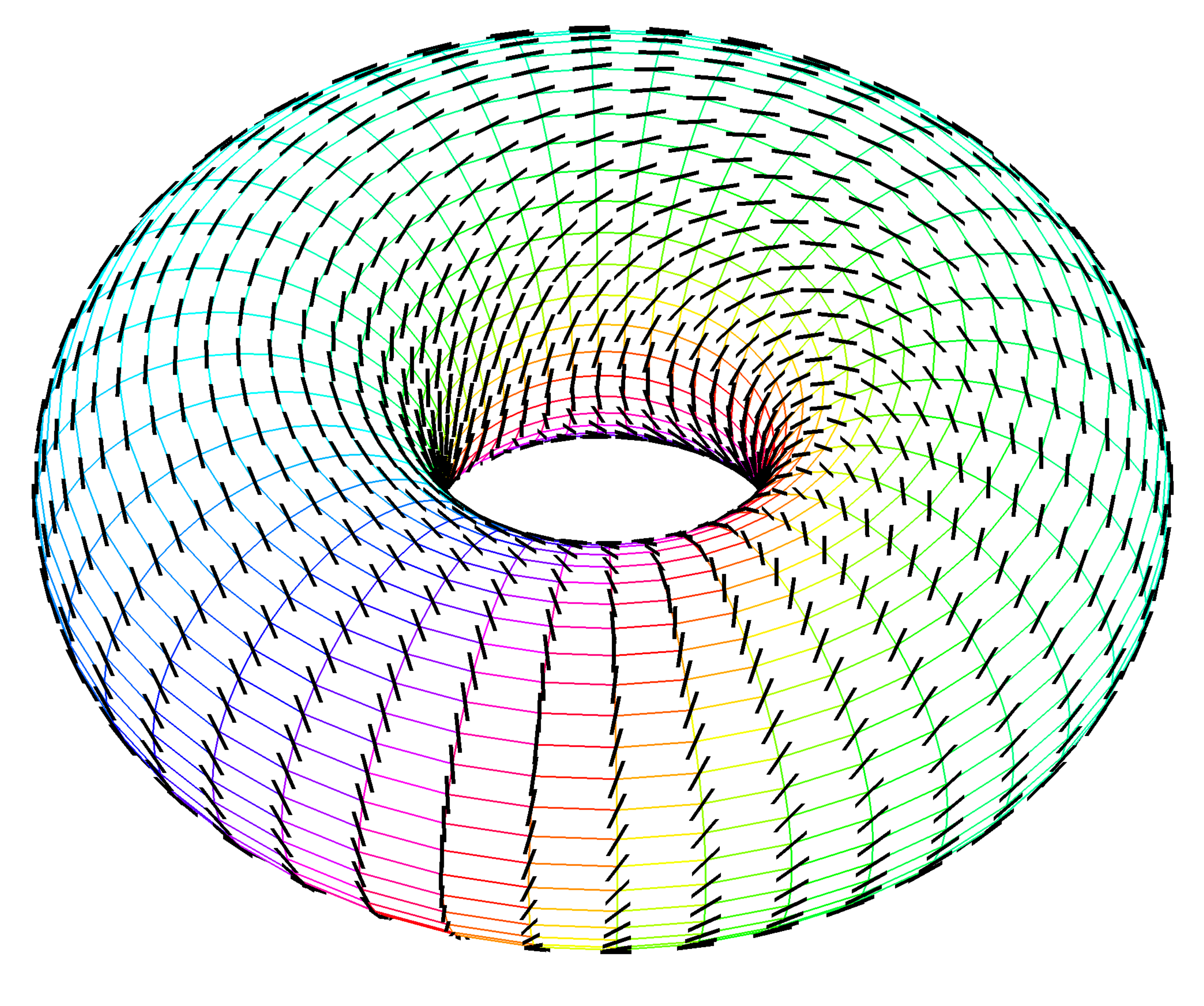} \qquad
				\labellist
			\hair 2pt
			\pinlabel $\mathbf{h}=(1,4),\quad\mbox{$W=17.15$}$ at 850 -35
		\endlabellist				
		 \includegraphics[height=3.5cm]{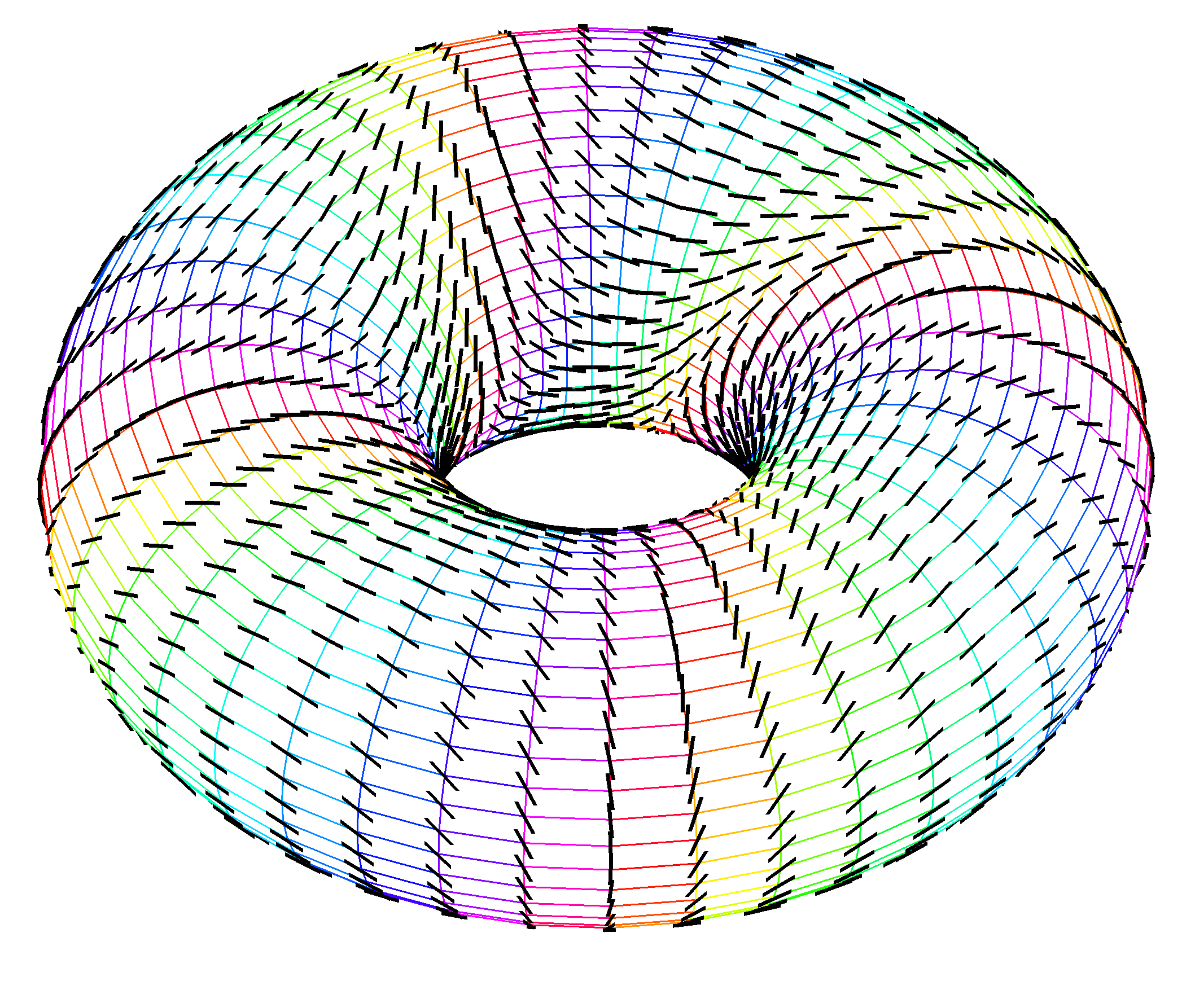} \qquad
		 \labellist
			\hair 2pt
			\pinlabel $\alpha$ at 1970 720
			\pinlabel $\mathbf{h}=(4,1),\quad\mbox{$W=23.02$}$ at 850 -35
		\endlabellist				
				\includegraphics[height=3.5cm]{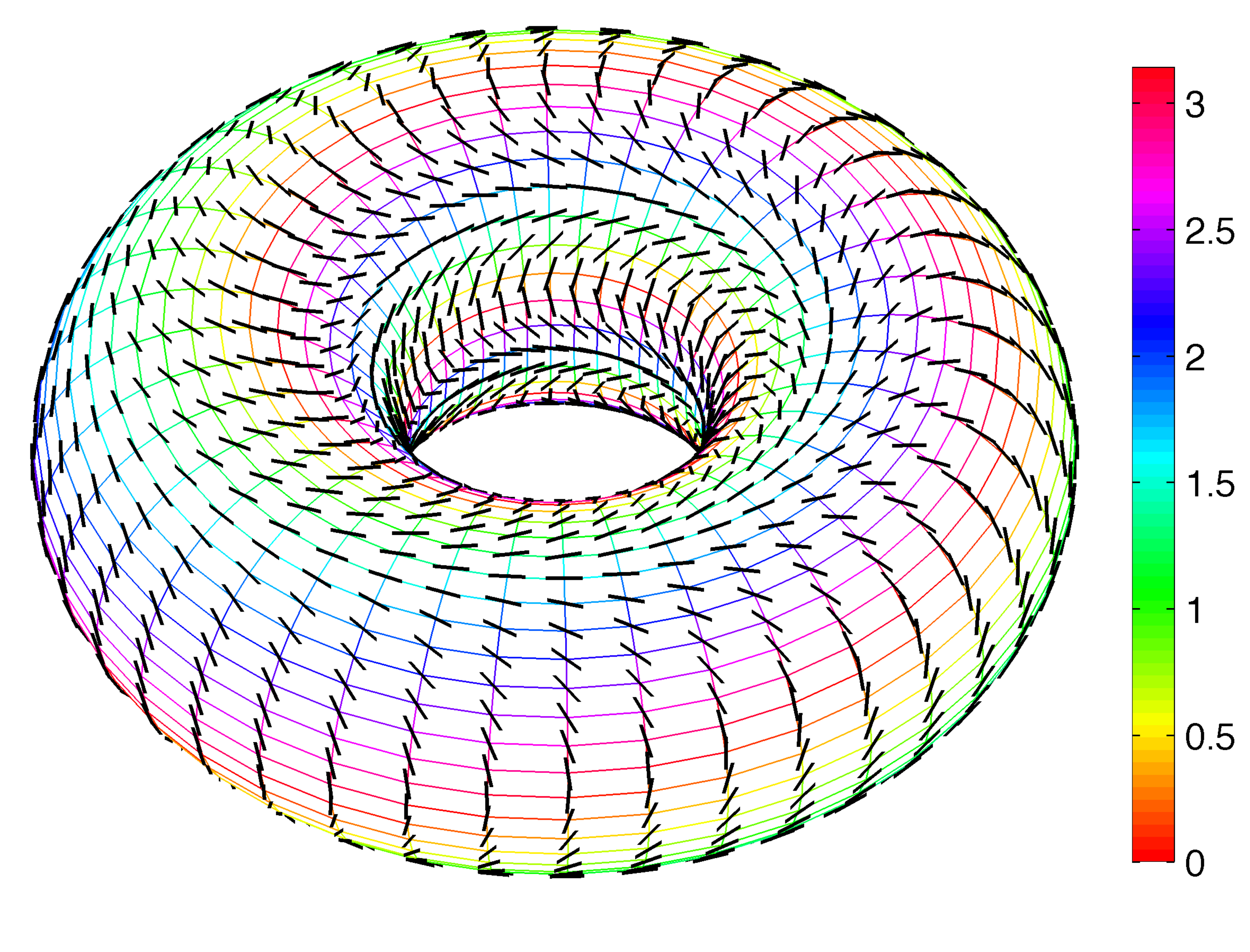} \\
		\caption{(Color online) Examples of (numerical) local minimizers with different winding numbers, on the torus with $\mu=1.8$. The unoriented segments represent the direction $\pm \n$. The winding number can be easily visualized by fixing a closed meridian or parallel line and counting how many times a complete (and oriented) scale of colors appears on the chosen line. The value $W$ under each figure indicates the value of the energy (rescaled by $k$) for that configuration. The energy value of the ground state $\alpha_p$, for this choice of parameters, is $W=9.85$.}
	\label{fig6}
\end{figure*}

Note that, by allowing for a difference of an odd multiple of $\pi$, we are
taking into account the symmetry $\n=-\n$ of the nematic represented by the
vector field $\n$. Geometrically, $h_\theta$ indicates how many turns of 180$^\circ$ 
are completed by $\n$ along the meridian line parametrized by $\theta 
\mapsto X(\theta,0)$; similarly, the number of turns along the parallel line 
$\phi \mapsto X(0,\phi)$ is given by $h_\phi$. A crucial property of the 
winding number is its invariance under continuous transformations of $\alpha$ 
(so that $\mathbf h$ could be equivalently computed on any 
pair of curves which are homotopically equivalent to the two that we 
chose). The relevant consequence is the following:  for any choice of $
\mathbf h=(h_\theta,h_\phi) \in \mathbb Z \times \mathbb Z$, there is 
an initial datum $\alpha_0$ (e.g., $\alpha_0(\theta,\phi)=h_\theta 
\theta +h_\phi \phi$) such that the corresponding vector field $\pm\n$ has 
winding number $\mathbf h$; the evolution of $\alpha_0$ according to 
\eqref{eq:gf} provides then, in the limit as $t\to +\infty$, a function $
\alpha_\infty$ whose associated line field $\n_\infty$ is a local 
minimizer of $W_\text{NV}$ and has winding number $\mathbf h$. 
Consequently, for each 
choice of winding number there is at least one configuration which is stable in the sense that it is not possible to lower its energy without breaking the pattern on a 
line, i.e., without creating a line defect. On the other hand,  large winding numbers impose large gradients on the director field, resulting in an increase in the intrinsic energy, monotone with respect to $|\mathbf h|$. In analogy with traditional elastic continua, it is reasonable to expect that a liquid crystal would break or yield after reaching a critical maximum distortion energy. We therefore expect a material-dependent upper bound on the observable winding number of the nematics field, which we hope could be determined experimentally.

\subsection{Defects on a torus}

Although defect-free ground states can be found on the torus
because of its zero Euler characteristic, these are not the only
minima  which can be found. Indeed, computational studies \cite{SelEtal2011}  using an
XY-model on tori of different radii ratios have found equilibrium
configurations with defects of positive and negative charges with net zero charge.  The reason we do not observe defects is not due to the choice parameters, but rather is inherent to the form of the energy $W_\text{NV}$. We claim that configurations with defects cannot be local minimizers of $W_\text{NV}$ \emph{unless the model is substantially modified}, because the energy associated to any defect is $+\infty$. 

We remark that our results are not in contradiction with those of \cite{SelEtal2011}. In fact, computing the energy of a defect on a discrete grid as in \cite{SelEtal2011} is equivalent to removing a disc centred at the defect with diameter equal to the distance between the grid nodes. We would expect that by refining the grid, the energy of the discrete configuration would tend to the energy of the continuous configuration. In particular, it would tend to $+\infty$ close to a defect. What prevents the formation of defects in our case is not a choice of the parameters $k_i$ or $\mu$, but rather the choice of the initial datum in the evolution law (6). In \cite{SelEtal2011} the initial configuration presents a random number of defects, which eventually pairwise annihilate, leaving survivors in multiples of 4 (or no defects at all). Instead, we choose defectless initial data, possibly with nontrivial winding number, which smoothly evolves into minimizers with the same winding.

The observations regarding defects are well-studied in the literature, and in particular they were anticipated by Lubensky and Prost in \cite{LubPro92} who introduced a ``short distance cutoff $r$".  By removing a small disc of radius $r$, they obtain an energy $\sim \log(1/r)$ corresponding to a ``condensation energy" in the small disc, or alternatively to the ``line tension" of the boundary created by removing the disc.  

\section{Concluding remarks}

We analyzed nematic liquid crystals on a toroidal particle, 
finding new equilibrium configurations for the surface energy recently 
proposed by Napoli and Vergori in \cite{NapVer12E,NapVer12L} and 
comparing them to the equilibria of classical energies. We identified a 
range of parameters where the new energy selects a unique equilibrium 
among the classical ones, and we showed the emergence of new 
equilibria configurations, in accordance with the added penalization of the 
normal curvature and geodesic torsion of the director field.

 We hope that 
experiments could be carried out in order to confirm our analysis.
\smallskip

\begin{acknowledgments}
The authors are grateful to F. Bonsante, G. Napoli, J.-C. Nave for 
inspiring conversations, remarks and suggestions. We also acknowledge E. G. Virga
for his constructive criticisms and suggestions on the manuscript. 
This research
started during a visit of A.S. to the Department of Mathematics and Statistics of McGill 
University, whose kind hospitality is acknowledged. M.S.
was introduced to the subject while supported by an NSERC USRA and is currently supported by an NSERC PGSM and FQRNT doctoral scholarships. 
Finally, A.S. and M.V. have been supported by the Gruppo Nazionale per l'Analisi 
Matematica, la Probabilit\`a e le loro Applicazioni (GNAMPA) of the Istituto
Nazionale di Alta Matematica (INdAM).
\end{acknowledgments}

\appendix

\section{Geometric quantities on the torus}
\label{app:torus}

Let $Q:=[0,2\pi]\times [0,2\pi] \subset \R^2$, and let $X:Q\to \R^3$ be the following parametrization of an embedded torus $\mathbb T$
\begin{equation}
\label{app:paramtorus}
	X(\theta,\phi) = 
		\begin{pmatrix} 
			(R+r\cos \theta)\cos \phi \\ 
			(R+ r\cos \theta)\sin \phi \\ 
			r\sin \theta
		\end{pmatrix}.
\end{equation}
Using parametrization \eqref{app:paramtorus},  in the next paragraph we derive the main geometrical quantities, like tangent and normal vectors, first and second fundamental form, in order to obtain an explicit expression for the metric and the curvatures of $\mathbb T$ and for $\nabla_s \n$.

Let $ X_\theta:= \partial_\theta X$, $X_\phi:= \partial_\phi X$, $\nbf:=\frac{X_\theta \wedge X_\phi}{|X_\theta \wedge X_\phi|},$
we have
\begin{gather*}
	X_\theta = \begin{pmatrix} 
			- r \sin \theta \cos \phi \\ 
			-r \sin \theta \sin \phi \\ 
			r \cos \theta 
			\end{pmatrix},\ \
	X_\phi = \begin{pmatrix} 
			-(R+r\cos \theta)\sin \phi \\ 
			(R+r\cos \theta)\cos \phi \\ 
			0 
			\end{pmatrix}, \displaybreak[2]\\
	X_{\theta\theta} = \begin{pmatrix} 
			-r\cos \theta \cos \phi \\ 
			-r\cos \theta \sin \phi \\ 
			-r \sin \theta 
			\end{pmatrix},\quad
	X_{\theta\phi} = \begin{pmatrix} 
			r\sin \theta\sin \phi \\ 
			-r\sin \theta \cos \phi \\ 
			0 
			\end{pmatrix}, \displaybreak[2]\\
	X_{\phi\phi} = \begin{pmatrix} 
			-(R+r\cos \theta)\cos \phi \\ 
			-(R+r\cos \theta)\sin \phi \\ 
			0 
			\end{pmatrix}, \quad
	\nbf = -\begin{pmatrix} 
			\cos \theta \cos \phi \\ 
			\cos \theta \sin \phi \\ 
			\sin \theta 
			\end{pmatrix}.
\end{gather*}
Note that this choice of tangent vectors yields an \emph{inner} unit normal $\nbf$.  The first and second fundamental forms are  
\[	
g = 
\begin{pmatrix} 
	r^2 & 0 \\ 
	0 & (R+r\cos\theta)^2 
\end{pmatrix},\qquad
	II = 
\begin{pmatrix} 
	\frac 1r  & 0\\ 
	0 & \frac{\cos\theta}{R+r\cos\theta} 
 \end{pmatrix}.
\]
We have $\sqrt g=r(R+r\cos\theta)$, $g^{ii}:=(g_{ii})^{-1}$. The principal curvatures are
\[
	c_1= \frac 1r,\qquad c_2=\frac{\cos\theta}{R+r\cos\theta}.
\]
The unit tangent vectors are
\[
\eu:= \frac{X_\theta}{|X_\theta|}=-
	\begin{pmatrix} 
		 \sin \theta \cos \phi \\ 
		 \sin \theta \sin \phi \\  
		\cos \theta 
	\end{pmatrix}\!\!,
	\ 
	\ed:= \frac{X_\phi}{|X_\phi|}= 
	\begin{pmatrix} 
		-\sin \phi \\ 
		\cos \phi \\ 
		0 
	\end{pmatrix}\!.
\]
The geodesic curvatures $\kappa_\theta$ and $\kappa_\phi$ of the principal lines of curvature can thus be obtained by
\begin{align*}
	\kappa_\theta =\ed(\nabla \eu)\eu &= \frac{1}{R+r\cos\phi}X_\phi \cdot \frac{1}{r^2} X_{\theta\theta} = 0,\\
\kappa_\phi = \ed(\nabla \eu)\ed &= \frac{X_\phi \cdot X_{\theta\phi}}{r(R+r\cos\theta)^2}  
		= \frac{-\sin\theta}{R+r\cos\theta}.
\end{align*}
The spin connection $\boldsymbol{\Omega}$ is the vector field with components given by
\begin{align*}
	\boldsymbol{\Omega}^1 &= (\eu,D_{\eu}\ed)_{\R^3} =-\kappa_\theta=0,\\
	\boldsymbol{\Omega}^2 &= (\eu,D_{\ed}\ed)_{\R^3} = -\kappa_\phi=\frac{\sin\theta}{R+r\cos\theta}.
\end{align*}
The explicit forms of the surface differential operators on the torus are
\begin{align*} 
	  \nabla_s \alpha &= g^{ii}\partial_i\alpha 
		= \frac{\partial_\theta \alpha}{r} \eu + \frac{\partial_\phi \alpha}{R+r\cos\theta} \ed,\displaybreak[2]   \\
	\Delta_s &=\frac{1}{\sqrt g}\partial_i(\sqrt g g^{ij}\partial_j)\displaybreak[2] \\
		&= \frac{1}{r^2}\partial^2_{\theta\theta} -\frac{\sin\theta}{r(R+r\cos\theta)}\partial_\theta 
			+ \frac{1}{(R+r\cos\theta)^2}\partial^2_{\phi\phi}.\displaybreak[2]
\end{align*}	
For $ \n = \cos\alpha \eu + \sin \alpha \ed$, the explicit expression of the surface gradient $\nabla_s \n$ in terms of the deviation angle $\alpha$, with respect to the Darboux frame $(\n,\tbf,\nbf)$ is
\begin{widetext}
\[
\nabla_s \n = 
\begin{pmatrix}
	0 & 0 & 0 \\
	\frac{\alpha_\theta}{r} \cos\alpha + \left(\frac{\alpha_\phi}{R+r\cos \theta} 
		- \frac{\sin\theta}{R+r\cos\theta}\right)\sin\alpha\quad
	& -\frac{\alpha_\theta}{r} \sin\alpha + \left(\frac{\alpha_\phi}{R+r\cos \theta} 
		- \frac{\sin\theta}{R+r\cos\theta}\right)\cos\alpha 	& 0 \\
	\frac{1}{r}\cos^2\alpha +  \frac{\cos\theta}{R+r\cos\theta}\sin^2\alpha
	&  \left( \frac{\cos\theta}{R+r\cos\theta}-\frac{1}{r}\right)\sin\alpha\, \cos\alpha  	& 0 
\end{pmatrix}.
\]
\end{widetext}

\section{Derivation of \eqref{eq:Walpha}}
The proof relies on algebraic manipulations and integration of trigonometric functions. Let $\mu:=R/r$, substituting the expressions for $c_1,c_2,\kappa_\phi,\sqrt g$ derived in Appendix A, we have
\begin{align*} 
	I_1:=\int_Q (\kappa_\phi)^2 \dvol &= \int_0^{2\pi}\int_0^{2\pi} \frac{\sin^2\theta}{\mu +\cos\theta}\d\theta \d\phi \\
			&= 4\pi^2 \left(\mu - \sqrt{\mu^2-1}\right),\\
	I_2:=\int_Q (c_1)^2 \dvol &= \int_0^{2\pi}\int_0^{2\pi} \left\{ \mu +\cos \theta \right\}\d\theta \d\phi = 4\pi^2 \mu,\displaybreak[2]\\
	I_3:=\int_Q (c_2)^2 \dvol &=  \int_0^{2\pi}\int_0^{2\pi} \frac{\cos^2\theta}{\mu +\cos\theta}\d\theta \d\phi \\
			& = 4\pi^2  \mu\left(\frac{\mu}{\sqrt{\mu^2-1}} -1\right),\displaybreak[2] \\
	\int_Q c_1c_2\, \dvol &= 0. 
\end{align*}
Ordering the terms according to the frequency in $\alpha$, we get
\begin{align*}
	W(\alpha) &=\left[ \frac{k_1 +k_3}{4}I_1 + \frac{k_2+k_3}{8}(I_2+I_3)\right] \displaybreak[2]\\
		&\quad + \cos(2\alpha)\left[ \frac{k_1-k_3}{4}I_1 + \frac{k_3}{4}(I_2-I_3)\right]\displaybreak[2]\\
		&\quad +\cos^2(2\alpha)\left[ \frac{k_3-k_2}{8}(I_2+I_3)\right],
\end{align*}
and substituting the values of $I_i$ yields \eqref{eq:Walpha}. We also note that the recurring value of $(I_2+I_3)/4$	corresponds to Willmore's functional on a torus (see \cite{Helfrich73}), i.e.
\[
	\mathcal{W}(\mathbb T)  =\int_Q \left(\frac{c_1+c_2}{2}\right)^2\dvol
= \frac{\pi^2\mu^2}{\sqrt{\mu^2-1}}. 	
\]
Willmore's functional can be interpreted as an elastic bending energy for unconstrained membranes that have a flat configuration at rest.

\end{document}